\documentclass[aps,prb,twocolumn,showpacs,superscriptaddress,article,floatfix]{revtex4}
\usepackage{graphicx}

\usepackage{amsmath}
\usepackage{natbib}
\usepackage{bm}
\usepackage[dvips]{color}

\begin{document}

\title{Electronic structure and electric-field gradients analysis in $CeIn_3$}
\date{\today}

\author{S. Jalali Asadabadi}
\email[Electronic address: ]{sjalali@phys.ui.ac.ir}
\affiliation{Department
of Physics, Faculty of Science, University of Isfahan (UI)\\
Hezar Gerib Avenue, Isfahan 81744, Iran} \affiliation{Research
Center for Nano Sciences and Nano Technology\\ University of
Isfahan (UI), Isfahan 81744, Iran}

\begin{abstract}
Electric field gradients (EFG's) were calculated for the $CeIn_3$
compound at both $^{115}In$ and $^{140}Ce$ sites. The
calculations were performed within the density functional theory
(DFT) using the augmented plane waves plus local orbital (APW+lo)
method employing the so-called LDA+U scheme. The $CeIn_3$ compound
were treated as nonmagnetic, ferromagnetic, and antiferromagnetic
cases. Our result shows that the calculated EFG's are dominated
at the $^{140}Ce$ site by the Ce-4f states. An approximately
linear relation is intuited between the main component of the
EFG's and total density of states (DOS) at Fermi level. The EFG's
from our LDA+U calculations are in better agreement with
experiment than previous EFG results, where appropriate
correlations had not been taken into account among 4f-electrons.
Our result indicates that correlations among 4f-electrons play an
important role in this compound and must be taken into account.
\end{abstract}

\maketitle

\section{Introduction}
\label{introd} Hyperfine interactions provide sensitive physical
quantities such as electric field gradients (EFG's), which can be
used to shed light experimentally\cite{Kohori1999,Kohori2000} and
theoretically\cite{Lali2001,Betsuyaku2004} into the electronic
states of materials. In this paper we have focused on the $CeIn_3$
as an interested system composed of strongly correlated
4f-electrons. It is a cubic heavy-fermion (HF) local moment
antiferromagnetic (LMAF) system at ambient pressure with a
N\'{e}el temperature of 10.1 K \cite{Lawrence1980}. This
concentrated Kondo compound exhibits\cite{Mathur1998} various
fascinating and unexpected physical properties. The various
properties of this heavy fermion originate from the fact that one
cannot assign a definite localization to the 4f-states,
irrespective of the applied conditions to the compound. The
unexpected physical behavior may be then attributed to the degree
of localization of the 4f-electrons, i.e. the positions of the
4f-density of states (4f-DOS) with respect to the Fermi level.
The position of the 4f-DOS demonstrates the degree of
hybridization between localized 4f and conduction bands. The EFG
quantity is extremely sensitive to the anisotropic charge
distributions of the core electrons\cite{Yu1991}, as well as to
the aspherical electron density distribution of valance
electrons\cite{Schwarz1990}, and as a result to the valance
electronic structure. The EFG, thereby, can serve as a powerful
gauge for measuring such a degree of localization.

J. Rusz et al.\cite{Rusz2005} very recently calculated Fermi
surfaces of the $CeIn_3$ regardless its antiferromagnetic
ordering. Their calculations were performed in the localized
extreme limit within the open core treatment \cite{Jalali2002,
Jalali2004}. On the other delocalized extreme limit two
individual groups\cite{Lali2001, Betsuyaku2004} calculated the
EFG at the $^{115}In$ site in this compound. The calculations of
the former\cite{Lali2001} and later\cite{Betsuyaku2004} groups
were performed, respectively, in the antiferromagnetic and
nonmagnetic phases employing a similar method of the
full-potential linearized augmented plane waves
(FP-LAPW)\cite{Singh2006}.

In this paper, we have examined whether one can, using an
intermediate way, improve the previous results of the localized
and delocalized limits. For this purpose, we have employed the
LDA+U scheme\cite{Anisimov1991,Anisimov1993,Czyyk1994} and then
calculated the EFG's at both $^{115}In$ and $^{140}Ce$ sites. The
more advanced method of augmented plane waves plus local orbital
(APW+lo)\cite{Sjstedt2000,Madsen2001} were used to linearize the
energies. According to our knowledge, this is a first-report of
the EFG calculations within the LDA+U scheme employing APW+lo
method for this compound. \textit{Our spin-polarized calculations
demonstrate non-zero EFG at $^{140}Ce$ site in the presence of
spin-orbit coupling}. The EFG's, according to our knowledge, had
not been previously calculated at the cubic $^{140}Ce$ site in
this compound. The cubic symmetry of the Ce site explains why the
EFG's were less significant to be previously reported. We have
emerged, nevertheless, a new result that \textit{the EFG's are
dominated at the $^{140}Ce$ site by 4f-states and not as usual by
p-states.} \textit{The goal of this work is to illustrate an
approximately linear relationship between the values of EFG and
density of states (DOS) at Fermi level ($E_F$), viz. $EFG \propto
DOS(E_F)$}. According to our knowledge, such a linear
relationship had not been previously observed. We also aim to
justify about the tendency of the 4f-electrons in the ground
state of the antiferromagnetic $CeIn_3$ compound to show their
degree of localization. We have also found that correlations
among 4f-electrons influence semicore states of 5p-Ce.

\section{Details of the Calculations}
\label{detail} All the calculations in this work were performed in
the frame work of the density functional
theory\cite{Hohenberg1964,Kohn1965} (DFT). We have taken the
generalized gradient approximation\cite{Perdew1996} (GGA) into
account for the exchange-correlation functional. We have employed
the full-potential augmented plane waves plus local orbital
(APW+lo) method\cite{Sjstedt2000,Madsen2001} as embodied in the
WIEN2k code\cite{Blaha1999}. The muffin-tin radii $(R_{MT})$ were
chosen to be 2.2 {\AA} and 2.8 {\AA} for the In and Ce atoms,
respectively. It has been allowed to be the 4s 4p 4f orbitals of
the Ce and 5d 6s orbitals of In in the valance states. The
expansion of the wave functions inside the spheres in lattice
harmonics and in the interstitial region in plane waves were cut
off by the maximum eigenvalue of $l_{max}=10$ and the
$R_{MT}K_{max}=7$, respectively. The cutoff for the Fourier
expansion of the charge density and potential was taken to be
$G_{\max }  = 16{\rm{ }}\sqrt {Ry}$. We used a mixing parameter
of 0.001 in the Broyden's scheme to reduce the probability of
occurrence of the spurious ghostbands. A mesh of 165 special k
points was taken in the irreducible wedge of the first Brillioun
zone, which corresponds to the grids of $18 \times 18 \times 18$
in the scheme of Monkhorst-Pack\cite{Monkhurst1976}. In order to
perform the calculations nearly in the same accuracy for the case
of magnetic super cell compared to the non-magnetic unit cell, the
mesh of k points was reduced to 121 corresponding to $9 \times 9
\times 9$ grids. We diagonalized spin-orbit coupling (SOC)
Hamiltonian in the space of scalar
relativistic\cite{Koelling1997} eigenstates using a
second-variational procedure\cite{MacDonald1980} imposing (111)
direction on the Ce magnetic moments within a cut-off energy of 3
Ry. In order to take into account strong correlations of 4f Ce
states, we have used the LDA+U
method\cite{Anisimov1991,Anisimov1993,Czyyk1994}. In one side, it
is generally believed that, the on-site Coulomb repulsion U
integral, as an input parameter for the LDA+U calculations,
depends on the system under study. This means that the U
parameter can vary for an atom, i.e. Ce in this work, from one
case to the others. Here, we have used a value of 6.2 eV for the
U parameter of Ce in $CeIn_3$. Fortunately, this value, U = 6.2
eV, had been before calculated\cite{Kioussis1996} for our own
case of $CeIn_3$. Furthermore, on the other hand, the value
obtained for the U parameter depends on the method of
calculations. Unfortunately, the calculations of the U value were
performed within the linear-muffin-tin-orbital (LMTO)
method\cite{Kioussis1996}. Although the method of our
calculations, i.e. APW+lo, is different from the LMTO, but we have
noticed that for another case of face-centered-cubic $\gamma-Ce$
the value of 6.1 eV could give satisfactory results in agreement
with experiment\cite{Shick2000}. The U=6.1 eV value of
fcc-$\gamma-Ce$ is not so far from the U=6.2 eV value of the
$CeIn_3$ compound. Therefore, we did neither more elaboration to
calculate the Coulomb repulsion U within our APW+lo method, nor
change this value to optimize it. Another input parameter for the
LDA+U calculation is the exchange integral J. Here, we have used
0.7 eV for the J value. This J parameter can be derived using the
values of 8.34 eV, 5.57 eV, and 4.12 eV for the $F^2 ,{\rm{ }}F^4
,{\rm{ }}F^6$ Slater integrals\cite{Shick2000}, respectively,
within the following expression, which is valid for the f
electrons:

\begin{eqnarray}
J = \frac{1}{3}\left( {\frac{2}{{15}}F^2  + \frac{1}{{11}}F^4  +
\frac{{50}}{{429}}F^6 } \right) \approx 0.69eV. \label{equ1}
\end{eqnarray}

Finally, since the factor of (U-J) appears in the total energy of
the LDA+U method instead of U and J individually, we set J to
zero, and let U be equal to the effective value\cite{Sawada1998}
of $U_{eff}$ = U - J = 5.5 eV.
\section{Chemical, Magnetic and Electronic Structures}
\label{crysta}
\subsection{Chemical Structure}
\label{sub-crysta} 
\begin{figure}
 \begin{center}
  \includegraphics[width=8cm,angle=0]{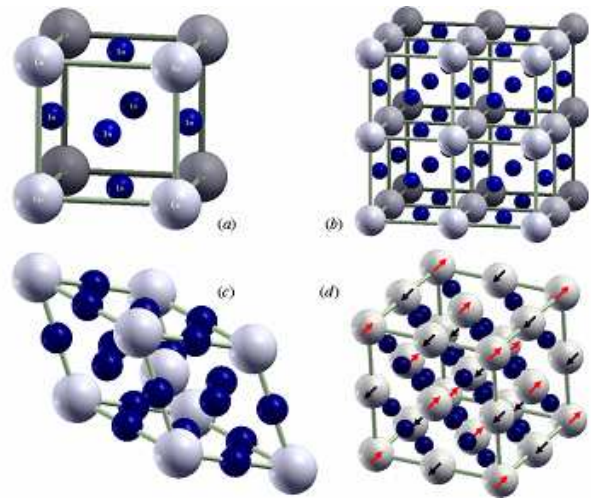}
   \caption{"(color online)"(a) Chemical unit cell of $CeIn_3$ in $AuCu_3$
prototype having the lattice parameter of a=4.69 A. (b)
Constructed conventional supercell with the symmetry of
face-centered cubic ($Pm\bar 3m$ space group) from the chemical
unit cell by a factor of 2 as a number of chemical unit cells
drawn in three directions of xyz Cartesian coordinates. (c)
Primitive rhombohedral unit cell of the constructed conventional
 supercell. (d) Magnetic supercell imposing spin ordering of $(\uparrow \downarrow)$
  to the Ce moments along (111) axis.
\label{fig1}}
 \end{center}
\end{figure}

As shown in Fig.~\ref{fig1}(a), $CeIn_3$
crystallizes in the space group of $Pm\bar 3m$ with the
binary-fcc prototype of $AuCu_3$. It is a cubic unit cell, where
Ce atoms are located on the corners, and the In atoms on the
middle of the surfaces. The point groups of Ce and In atoms are
the cubic m-3m, and the non-cubic 4/mmm, respectively. The
lattice parameter of the $CeIn_3$ were measured\cite{Harris1965}
to be 4.69 {\AA}. We have used this chemical structure to simulate
the nonmagnetic and ferromagnetic phases.

\subsection{Magnetic Structure}
\label{sub-magnet} It has been experimentally observed that the
cerium moments in this compound are aligned antiferromagnetically
in adjacent (111) magnetic planes \cite{Lawrence1980}. Therefore,
in order to simulate the antiferromagnetic situation, first the
sides of the non-magnetic unit cell were doubled as depicted in
Fig.~\ref{fig1}(b) in all the 3 Cartesian xyz directions. We have
preserved the space group of $Pm\bar 3m$ for the new magnetic
supercell. The preserved fcc symmetry of the magnetic supercell
causes to be reduced the number of atoms in the primitive
rhombohedral magnetic unit cell, see Fig.~\ref{fig1}(c), compared
to the conventional unit cell shown in Fig.~\ref{fig1}(b). Second,
we have imposed the antiferromagnetic ordering on the magnetic
moments of $^{140}Ce$ sites. The direction of the 4f spin in the
ground state electron configuration of Ce atom, i.e. $[Xe].4f^
\uparrow  .5d^ \uparrow .6s^{ \uparrow \downarrow } $, were
exchanged, i.e. $[Xe].4f^ \downarrow  .5d^ \uparrow  .6s^{
\uparrow \downarrow } $, alternatively with the ordering of $(
\uparrow  \downarrow )$ in the (111) direction as shown in
Fig.~\ref{fig1}(d).

\subsection{Electronic Structure}
\label{sub-electr} We have calculated the density of states (DOS)
in the lack of both spin-polarization (SP) and spin orbit coupling
(SOC). This calculation has been performed using the chemical
structure; Fig.~\ref{fig1}(a). This is what we call it
nonmagnetic (NM) phase from later on. The SOC was then included in
two individual steps. The calculations in the first step were
performed including spin-orbit interactions among only
Ce-electrons. We refer to the results of this calculation by the
name of "NM+SOC(only Ce)". The SOC was then, as the second step,
included among the In-electrons as well. We call it "NM+SOC"
phase form now on. For sure, the SOC is included in both Ce and In
atoms in the "NM+SOC" phase. Our calculated DOS's are in
agreement with the previous calculations\cite{Lali2001}, and we
avoid repeating them here. The SOC not only influences the
semicore states of 5P Ce and 4d In, but also changes the density
of states at the Fermi level ($DOS(E_F)$). The SOC splitting in
the valance states is much smaller than the SOC splitting in the
semi core states. However, the effects of the SOC would not be
neglected on the results due to the fact that many physical
quantities, e.g. EFG, are very sensitive to the value of the
$DOS(E_F)$. The later point seems to be more significant for the
Ce based compounds. A typical LDA/GGA calculation produces a
sharp density of states for the 4f Ce, and situates it at the
Fermi level. Thereby, any small changes in the sharply located 4f
Ce DOS at the Fermi level can change more significantly the
$DOS(E_F)$, and consequently the physical results.

\begin{figure*}
 \begin{center}
  \includegraphics[width=18cm,angle=0]{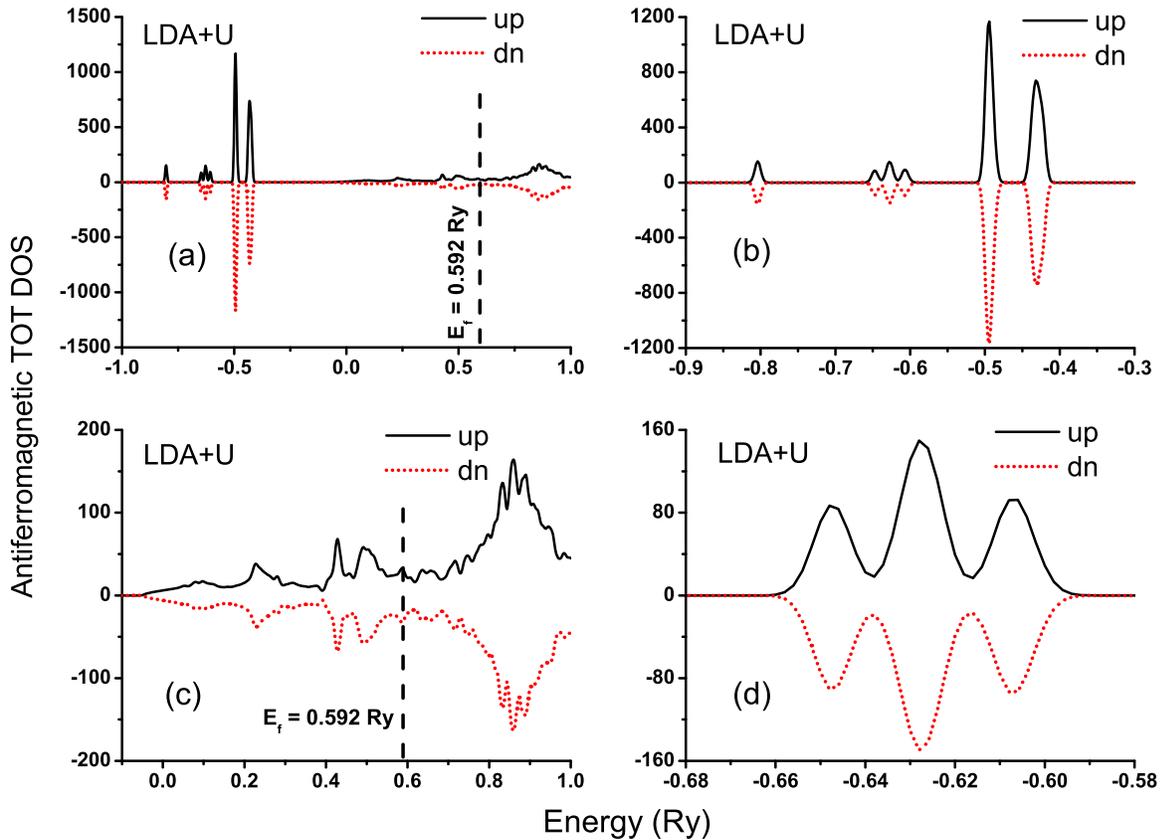}
  \caption{"(color online)" Total density of states of the AFM(111)+SOC+LDA+U
  phase presenting (a) semicore and valance states, b) semicore states,
(c) valance states, (d) selected interval of semicore states
exhibiting further splitting in 5p Ce DOS from LDA+U interaction.
Fermi level is shown using dashed lines
 \label{fig2}}
 \end{center}
\end{figure*}

\begin{figure}
 \begin{center}
  \includegraphics[width=9cm,angle=0]{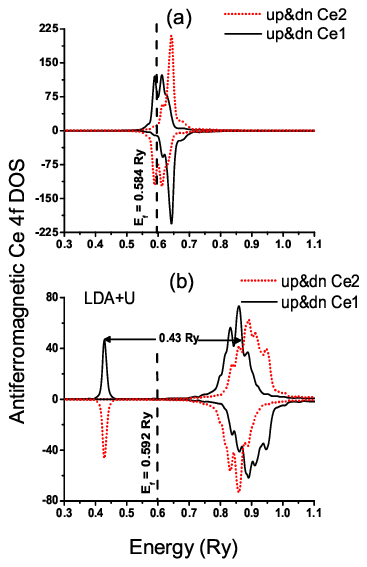}
  \caption{"(color online)" 4f-DOS of the (a) AFM(111)+SOC phase and (b) AFM(111)+SOC+LDA+U
  phase. Solid (dotted) curves show up and down DOS's
of the Ce1 (Ce2), where label 1 (2) refers to the attributed $
\uparrow$ ($\downarrow$) ordering. Fermi levels are shown using
dashed lines for each of cases.
 \label{fig3}}
 \end{center}
\end{figure}

The DOS's were also calculated for the ferromagnetic phase in the
lack of SOC, which we call it "FM" phase. The result shows that
imposing spin polarization (SP) causes to be reduced the
$DOS(E_F)$ from 141.82 states/(spin.Ry) in NM phase to 94.46
states/(spin.Ry) for spin up and to 14.90 states/(spin.Ry) for
spin down in the FM phase. We have added up and down DOS's of the
FM phase. Therefore, a total reduction of 32.46 states/Ry in
$DOS(E_F)$ is occurred in going from NM to FM phase. Such a
reduction affects the physical quantities that we shall discuss
them in subsequent sections. We have then included the SOC in the
FM phase, which we call it "FM+SOC" phase. The result shows that
including SOC gives rise to a reduction of 14.60 states/Ry from
 FM to FM+SOC. The $DOS(E_F)$ shows a total reduction of 47.06
states/Ry in going from NM to FM+SOC. Hence it seems that the SP
and SOC can influence physical quantities in a similar direction
in this compound.

The DOS's were calculated for the AFM state in the lack of SOC.
The later state is called "AFM(001)" phase in this paper. In the
absence of the SOC there is no preferred spatial direction at the
cubic $^{140}Ce$ site. In this case all the planes, for example
(111), (001) and so on, are identical. Thus, for simplicity, the
antiferromagnetic ordering, $( \uparrow \downarrow )$, were
aligned along (001) axis. The SOC interactions were then included
in the AFM(001) phase. One could introduce, at this step, a
preferable direction in the presence of SOC. However, we would
postpone setting up the correct direction of the cerium moments.
The cerium moments were then kept still along (001) axis, which we
call it "AFM(001)+SOC" phase. The cerium moments in the
AFM+SOC(001) phase were not changed to the more natural (111)
direction, in order to avoid mixing the effects of SOC with the
effects of spin directions in the AFM phase. The result of
AFM(001) or AFM(001)+SOC shows that the picks of the semicore Ce
5p and 4d In DOS's were nearly doubled in the AFM compared to the
NM and FM phases. This is in consistent with the fact that the
number of electrons are doubled in going from the chemical cell
to the magnetic supercell. We have then directed the 4f Ce spins
from (001) to (111) direction, and recalculated the density of
states self-consistently including spin-orbit coupling. The DOS's
qualitatively were similar to those for the last case of (001)
direction. However, quantitatively only changing the direction
changes the values of $DOS(E_F)$ from 167.47 sates/(spin.Ry) to
127.98 sates/(spin.Ry) for the spin up, and from 167.24
sates/(Ry.spin) to 128.24 states/(spin.Ry)  for spin down. Such
changes can affect the sensitive physical quantities.

We have included an appropriate correlation among 4f Ce electrons
with employing LDA+U calculations. The 4f spin orientations were
preserved antiferromagnetically along (111) direction, and
spin-orbit interactions were included in the LDA+U calculations.
This constitutes our "AFM(111)+SOC+LDA+U" phase. Total DOS's of
the AFM(111)+SOC+LDA+U phase are shown in Figs.~\ref{fig2}. The
result shows a significant reduction of the $DOS(E_F)$ to 28.11
states/(spin.Ry) for both up and down spins. Total DOS's are
separately illustrated in semicore, Fig.~\ref{fig2}(b), and
valance, Fig.~\ref{fig2}(c), regions. The DOS of the semicore
region shows further splitting in one of the branches of 5p Ce
DOS. This occurs in comparison with the phase of AFM(111)+SOC.
Our calculated total DOS of the later phase, which is not shown
here, is in complete accord with previous
calculations\cite{Lali2001}. One may confirm the above mentioned
further splitting by comparing Fig. 1 given in Ref.
\onlinecite{Lali2001} with our Fig.~\ref{fig2}(a) or (b). One
would more clearly observe the further splitting in
Fig.~\ref{fig2}(d). The semicore region in Fig.~\ref{fig2}(d) is
restricted to an appropriate interval, where the splitting is
occurred. This shows that \textit{correlations among 4f electrons
influence not only 4f Ce states directly, but also 5p Ce semicore
states indirectly in this compound}. There are two nonequivalent
Ce atoms in the magnetic cell with opposite 4f spin directions,
which are indexed as Ce1 and Ce2. The 4f DOS's of Ce1 and Ce2 are
shown in Figs.~\ref{fig3} for both phases of AFM(111)+SOC and
AFM(111)+SOC+LDA+U. The up and down 4f-DOS's of Ce1 are
asymmetric with respect to each other, which is the case for Ce2
as well. This is in the case that in overall Ce 4f DOS is
entirely symmetric regardless indexes 1 and 2. \textit{Therefore,
in spite of the fact that Ce1 and Ce2 can individually impose
magnetic moment, no net magnetic moment is imposed in the whole
of the supercell.} The 4f DOS of the AFM(111)+SOC phase is also
shown in Fig.~\ref{fig3} (a). In this case one can more easily
compare it with the later phase of AFM(111)+SOC+LDA+U shown in
Fig.~\ref{fig3} (b). The result shows that the significant
reduction in the $DOS(E_F)$ originates mainly from the splitting
between occupied and unoccupied 4f bands; see Fig.~\ref{fig3}
(b). Such a splitting within LDA+U treatment is due to the
effective Coulomb repulsion Hubbard U parameter. The 4f Ce DOS
piles up in the vicinity of Fermi level within our GGA
calculations. The later point is what one can clearly see for the
AFM(111)+SOC phase in Fig.~\ref{fig3}(a). This is in the case
that the LDA+U calculations give rise to be constituted the 4f Ce
DOS's far from Fermi level. As shown in Fig.~\ref{fig3}(b),
including repulsion U potential energy causes to be shifted down
occupied and shifted up unoccupied 4f DOS's with respect to Fermi
level. The separation energy between upper and lower 4f bands is
calculated to be 0.43 Ry. Thus contributions of 4f Ce to the
value of $DOS(E_F)$ are highly decreased. The receding of the 4f
Ce DOS, from the vicinity of Fermi level, indicates that the 4f
electrons are going to lose their itinerant character, and as a
result to gain their localized character. In order to justify
about the tendency of the 4f-electrons to clarify the degree of
their localization, one can compare Fig.~\ref{fig2}(c) with
Fig.~\ref{fig3}(b). The comparison shows that \textit{the 4f Ce
states play an important role and must not be ignored}. The later
point may be deduced from two subsequent facts. First in energy
space, the 4f states as shown in Fig.~\ref{fig3}(b) are
distributed over an energy interval for which other conduction
bands, e.g. In-states, as shown in Fig.~\ref{fig2}(c) are
distributed as well. Second in real space, each In atom is
surrounded by 4 Ce atoms as one can see in Fig.~\ref{fig1}(a).
Consequently, the 4f states are well hybridized with the other
conduction bands even after including LDA+U. Thus, the
localization is reduced but not vanished by LDA+U. Therefore, we
conclude that \textit{not only band-like treatment, but also
open-core treatment cannot provide satisfactory results for the
case of localized 4f states in this CeIn3 compound.} The former
treatment would not be well trusted because of putting 4f states
right at the Fermi level. Thus the band-like treatment
overestimates hybridization of the 4f Ce states with the other
valance states. The later treatment would not be also well
trusted because of confining 4f states in the core region. Thus
the open-core treatment usually underestimates hybridization of
4f Ce states with the other valance states.

\section{electronic specific heat}
\label{cv}

\begin{figure}
 \begin{center}
  \includegraphics[width=11cm,angle=0]{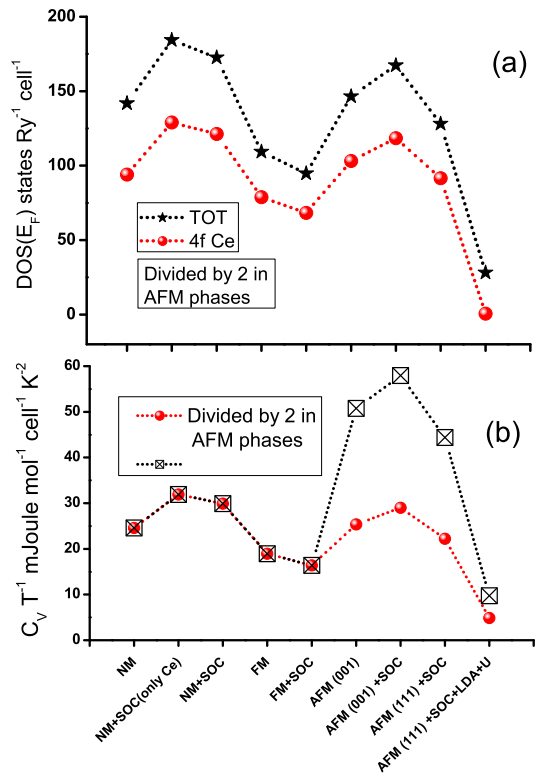}
  \caption{"(color online)" (a) Total and 4f-Ce DOS's versus a variety of phases discussed in
the text at their respective Fermi levels.  The values of
DOS(E$_{F})$ are divided by 2 for all the AFM phases. (b)
Sommerfeld linear coefficient $\gamma  = C_V /T$ in $mJoule.mol^{
- 1} .cell^{-1}.K^{ - 2}$ of the electronic specific heat shown
by crossed-squares versus all the phases. For comparison the
$\gamma$ coefficients, as shown by filled circles, are divided by
2 for the AFM phases.
 \label{fig4}}
 \end{center}
\end{figure}

In this section, to fix ideas it seems advisable to compare the
behavior of the electronic specific heats with the behavior of
the $DOS(E_F)$ through introduced phases in the preceding
sec.~\ref{sub-electr}. \textit{The findings of this comparison in
the next sec.~\ref{electr} will be of relevance to the goal of
this paper to realize whether or not such a comparison can be
generalized to the EFG quantity as well.} Therefore, we have
plotted in Fig.~\ref{fig4}(a) the calculated total and 4f Ce
$DOS's(E_F)$ versus the discussed phases. The number of atoms in
all the AFM phases is two times the number of other phases. The
later point is also the case for the $DOS's(E_F)$. We have then
divided the $DOS's(E_F)$ by 2 for all the AFM phases. In this
case, one can compare the AFM-$DOS's(E_F)$ with the $DOS's(E_F)$
of the other phases regardless the number of atoms in the
magnetic and chemical cells. One performing such a comparison can
focus only on the magnetic ordering effects through all the
defined various phases. \textit{The behavior of this curve
constitutes the backbone of this paper}.

We have calculated the electronic specific heats, $C_V$, in the
lake of both phonon-phonon and electron-phonon interactions, and
plotted the
 Sommerfeld linear coefficient, $\gamma = C_V/T$, in Fig.~\ref{fig4}(b) versus all the
 phases. The result nicely represent the behavior of
the total $DOS(E_F)$ shown in Fig.~\ref{fig4}(a), provided that
the calculated specific heats per cell were also divided by 2 for
the AFM phases. One would not be surprised observing such a
perfect consistency between in one side the behavior of the
specific heats and on the other hand the behavior of $DOS's(E_F)$
going through our defined phases. One would not be so, because
the specific heats calculations were performed taking only
electron-electron interactions into account. One can analytically
omitting phonon interactions easily prove\cite{Kittle2005} the
formula of $C_V /T = 1/3\pi ^2 k_B^2 DOS(E_F )$. This formula
demonstrates a linear relation between $\gamma$ and $DOS(E_F)$.
According to our knowledge, \textit{it had not been analytically
established such a relation between electric filed gradients
(EFG's) and $DOS(E_F)$.} Therefore, the above sketched strategy
will numerically make an opportunity in the next section trying to
demonstrate an approximately linear relation between EFG and
$DOS(E_F)$. We close this section by reporting the value of 9.74
$mJ/(mol.cell.K^2)$ within our LDA+U calculations. This calculated
$\gamma$ value is almost one order of magnitude less than the
experimentally measured \cite{Nasu1971} value of $130{\rm{
mJoule/(mol}}{\rm{.cell}}{\rm{.K}}^{\rm{2}} {\rm{)}}$. The
discrepancy is in agreement with all the other ab initio
calculations for other cases in the lack of phonon
interactions\cite{Petit2003, Albers1986, Steiner1994}.

\section{Electric Field Gradient}
\label{electr}

\begin{table*}
 \begin{center}
 \caption{The main component of the EFG, $V_{zz}$, and its
decomposition to valance, $V_{zz}^{val}$, and lattice,
$V_{zz}^{lat}$, components given in the units of $10^{21} V/m^2$
at both In and $^{140}Ce$ sites for all the phases discussed in
the text together with the calculated results within LAPW method
by the others as well as experimental EFG. \label{tab1}}
  \begin{ruledtabular}
   \begin{tabular}{lccccccccccccccc}
&  & Inte.&& && $V_{zz}$ & & & $V_{zz}^{val}$ & & &
$V_{zz}^{lat}$  \\  \cline{2-14}
Phase & SOC && LDA+U&$\mathop{\overrightarrow M ^{||} }\limits^{}$& In & & Ce & In  & & Ce & In && Ce  \\
      \hline
   \hspace{.55cm}NM & No && No      && 13.209 &  & 0 & 13.263 &  & 0 & -0.054 &  &     0  \\
   \hspace{.55cm}NM & Only Ce && No &&12.466 &  & 0 & 12.515 &  & 0 & -0.049 &     & 0  \\
   \hspace{.55cm}NM & Yes&& No &&12.847 &  & 0 & 12.538 &  & 0 & -0.051 &    & 0  \\
   \hspace{.55cm}FM & No      && No &&13.018 &  & 0 & 13.024 &  & 0 & -0.006 &     & 0  \\
   \hspace{.55cm}FM & Yes     && No &(001)&12.843 &  & 0.082 & 12.867 &  & 0.105 & -0.024 &  & -0.023  \\
   \hspace{.3cm}AFM & No  && No &(001)& 13.027 &  & 0.089 & 13.032 &  & 0.086 & -0.005 &   & -0.003 \\
   \hspace{.3cm}AFM & Yes && No &(001)& 12.966 &  & 0.086 & 12.968 &  & 0.066 & -0.002 &  & -0.002 \\
   \hspace{.3cm}AFM & Yes && No &(111)& 12.857 &  & 0.029 & 12.961 &  & 0.040 & -0.004 &   & -0.011 \\
   \hspace{.3cm}AFM & Yes && Yes &(111)& 12.431 &  & -2.863 & 12.442 &  & -2.889 & -0.011 && 0.026 \\ \hline\hline
   $^{\odot}AFM$ & Yes && No &(111)& 12.490 &  & - & 12.540 &  & - & -0.050 &   & - \\ \hline
   $^{\otimes}AFM^{exp}$ &  &   &&unknown& 11.6 &  & - & - &  & - & - &    & -

    \end{tabular}
  \end{ruledtabular}
 \end{center}
\begin{flushleft}
$\left\{ \begin{array}{l}
^{\odot}Ref.~\textrm{\onlinecite{Lali2001}} \\
^{\otimes}Ref.~\textrm{\onlinecite{Kohori1999}} \\
\end{array} \right.$

\end{flushleft}

\end{table*}

\begin{table*}
 \begin{center}
 \caption{Valance p and d anisotropy functions, $\Delta p(E_F)$ and $\Delta d(E_F)$,
  evaluated at In and $^{140}Ce$ sites for the
discussed phases in the text.
 \label{tab2}}
  \begin{ruledtabular}
   \begin{tabular}{lccccccccccc}
&  & Inte.&& &&$\Delta p$ & & & $\Delta d$ \\
 \cline{2-11}
     Phase & SOC && LDA+U&$\mathop{\overrightarrow M ^{||} }\limits^{}$& In & & Ce & In            & & Ce  \\
      \hline
   \hspace{.25cm}NM & No && No      &&0.1066 &  & 0 & -0.0060 &  & 0  \\
   \hspace{.25cm}NM & Only Ce && No &&0.1040 &  & 0 & -0.0061 &  & 0   \\
   \hspace{.25cm}NM & Yes     && No &&0.1043 &  & 0 & -0.0061 &  & 0    \\
   \hspace{.25cm}FM & No      && No &&0.0373 &  & 0 & -0.0050 &  & 0     \\
   \hspace{.25cm}FM & Yes     && No &(001)&0.0369 &  & 0 & -0.0052&  & 0  \\
   AFM & No  && No &(001)&0.0373 &  & 0.00015 & 0.0025&  & 0.00130   \\
   AFM & Yes && No &(001)&0.0372 &  & 0.00005 & 0.0024 &  & 0.00130   \\
   AFM & Yes && No &(111)&0.0370 &  & -0.00010 & 0.0024 &  & -0.00085   \\
   AFM & Yes && Yes &(111)& 0.0361 &  & 0.00715 & 0.0026 &  &
   0.01880
    \end{tabular}
  \end{ruledtabular}
 \end{center}
\end{table*}

\begin{figure*}
 \begin{center}
  \includegraphics[width=18cm,angle=0]{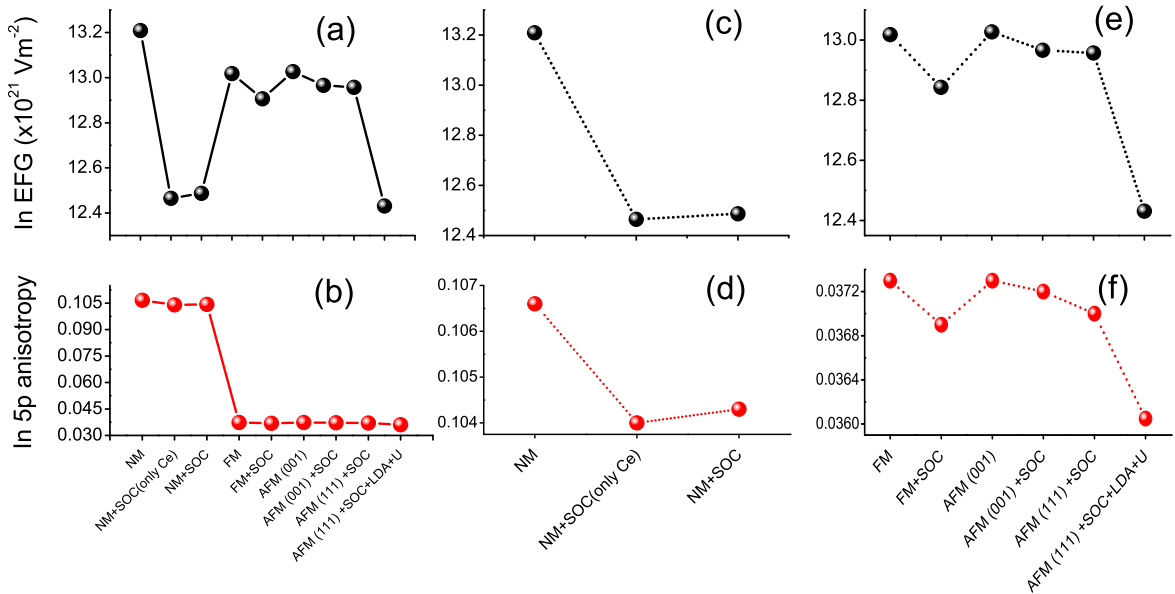}
   \caption{"(color online)" (a) The calculated EFG at the $^{115}In$ site and (b)
   the In 5p anisotropy function, $\Delta p(E_F)$, for all the nonmagnetic and magnetic
phases discussed in the text, which they are also represented in
(c) and (d) for only nonmagnetic phases as well as in (e) and (f)
for only magnetic phases.
  \label{fig5}}
 \end{center}
\end{figure*}

\begin{figure}
 \begin{center}
  \includegraphics[width=9cm,angle=0]{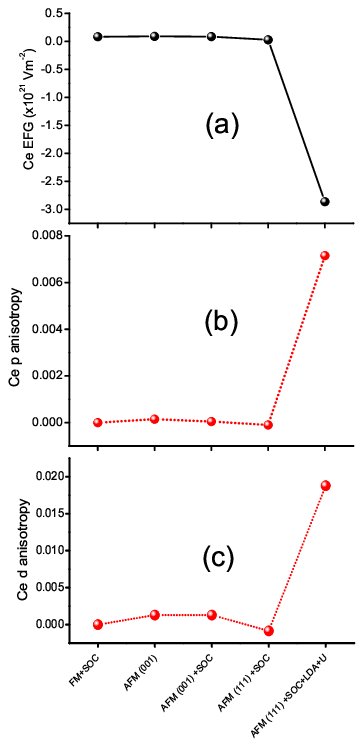}
  \caption{"(color online)" (a) The calculated EFG at the $^{140}Ce$ site, (b) the Ce p anisotropy
function, $\Delta p(E_F)$, and (c) the Ce d anisotropy function,
$\Delta d(E_F)$, for the phases discussed in the text for which
nonzero EFG's have been calculated at the $^{140}Ce$ site, viz.
ranging from FM+SOC to AFM(111)+SOC+LDA+U.
 \label{fig6}}
 \end{center}
\end{figure}

The electric field gradient (EFG) is a tensor of rank 2. The EFG
tensor has only 2 independent components in the principle axes
system (PAS). The axes of the system were chosen such that
$\left| {V_{zz} } \right| \ge \left| {V_{yy} } \right| \ge \left|
{V_{xx} } \right|$. One can then only evaluate the main $V_{zz}$
component of the EFG and the asymmetry parameter $\eta =
\frac{{V_{xx}  - V_{yy} }}{{V_{zz} }}$ to determine the two
independent components of the EFG in the PAS. In this paper we
only focus on the $V_{zz}$ as our calculated electric field
gradients, since the asymmetry parameters are zero for our case.
The main component of the EFG tensor has been calculated using
the following formula\cite{Blaha1988}:
\begin{eqnarray}
V_{zz}  = \lim _{r \to 0} \sqrt {\frac{5}{{4\pi }}} \frac{{V_{20}
}}{{r^2 }}, \label{equ2}
\end{eqnarray}
where radial potential coefficient, $V_{20}$, within LAPW method
has been calculated as follows\cite{Blaha1988,Schwarz1990}:
\begin{eqnarray}
V_{20} (r = 0) &=& \frac{1}{5}\int\limits_0^{R_{MT} } {\frac{{\rho
_{20} }}{{r^3 }}\left\{ {1 - \left( {\frac{r}{{R_{MT} }}}
\right)^5 } \right\}d^3 r}\nonumber\\
 &+& 4\pi \sum\limits_K
{V(K)j_2 (KR_{MT} )Y_{20} (\hat K)}. \label{equ3}
\end{eqnarray}

The integral yields the EFG contribution of the electrons
 inside and over the surface of the muffin-thin sphere
with radius of $R_{MT}$. The summation yields the EFG
contribution of the electrons entirely outside of the spheres.
The contribution of the electrons inside the sphere is called the
valance EFG, which we here denote it by $V_{zz}^{val}$. The
contribution of the electrons over the surface and outside of the
spheres is called the lattice EFG, which we here denote it by
$V_{zz}^{lat}$.

\begin{figure}
 \begin{center}
  \includegraphics[width=8cm, angle=0]{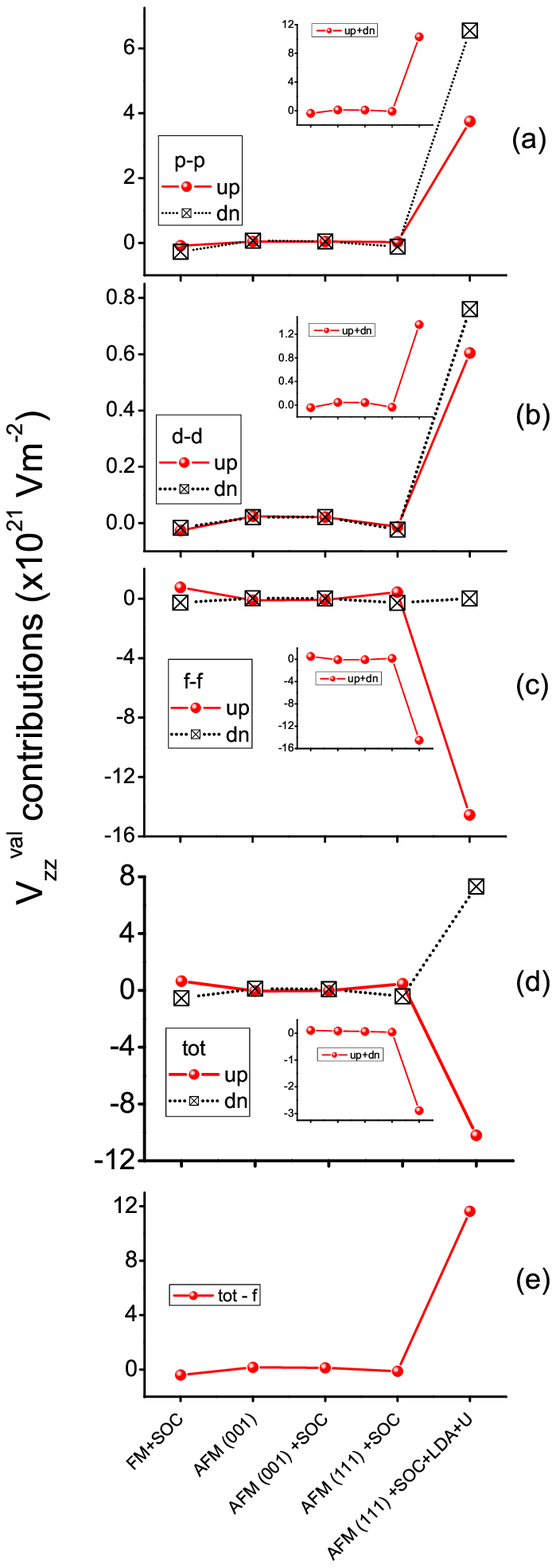}
   \caption{"(color online)" (a) Decomposed up, $\uparrow$, down, $\downarrow$, and
up plus down, $\uparrow + \downarrow$(see insets), valance
contributions of EFG evaluated at $^{140}Ce$ site for the
magnetic phases discussed in the text ranging from FM+SOC to
AFM(111)+SOC+LDA+U. (NB: There are no any EFG contributions at Ce
site for FM phase, in the lack of SOC, due its cubic symmetry.)
  \label{fig7}}
 \end{center}
\end{figure}

Electric field gradients, $V_{zz}$'s, and their respective
valance, $V_{zz}^{val}$'s, and lattice, $V_{zz}^{lat}$'s,
components were calculated for all the introduced phases in
sec.~\ref{sub-electr}. Our results are compared with experiment
and other theoretical calculations in Tab.~\ref{tab1}. The
anisotropy functions of $\Delta p (E_F)$ and $\Delta d (E_F)$
were also calculated and listed in Tab.~\ref{tab2} for all the
phases. \textit{The result, in Tab.~\ref{tab1}, shows that our
calculated EFG within the phase of AFM(111)+LDA+U is in better
agreement with experiment than previous
calculations\cite{Kohori1999, Lali2001}.} The better agreement
confirms that we have taken more properly correlations among 4f
electrons into account within our LDA+U calculations. One
observes, from Tab.~\ref{tab2}, that at In site the absolute
value of $\Delta p (E_F)$ is one order of magnitude greater than
the absolute value of $\Delta d (E_F)$, which is not the case at
Ce site. For the later case, the absolute value of $\Delta p
(E_F)$ is one order of magnitude smaller than the absolute value
of $\Delta d (E_F)$, i.e. $|\Delta p (E_F)|<|\Delta d (E_F)|$. We
will back to this point soon. What important for us here is to
study the variation of the EFG and appropriate anisotropy
functions versus discussed phases to obtain a relation between
EFG and $DOS(E_F)$. Therefore, we perform more comparisons
throughout Tab.~\ref{tab1} and Tab.~\ref{tab2} illustratively in
Figs.~\ref{fig5}. We have plotted $V_{zz}$ in Fig.~\ref{fig5}(a)
and $\Delta p$ in Fig.~\ref{fig5}(b), both evaluated at
$^{115}In$ site, versus all the phases. It is generally believed
that, \textit{contributions to the EFG originating from p states
dominate \cite{Blaha1988}.} Consequently, one expects to find a
linear relation between EFG and $\Delta p$, i.e.
$EFG\propto\Delta p$. This is what one expects to observe looking
at Fig.~\ref{fig5}(a) and Fig.~\ref{fig5}(b). It is hard to
realize, however, the linear relation of $EFG\propto\Delta p$ by
comparing these two figures with each other. Therefore, in order
to exhibit such a linear relationship, we have separated 3
nonmagnetic phases from the other 6 magnetic phases. The $V_{zz}$
and $\Delta p$ were respectively shown in Fig.~\ref{fig5}(c) and
Fig.~\ref{fig5}(d) versus 3 nonmagnetic phases. Similar results,
$V_{zz}$ and $\Delta p$, for the magnetic phases were separately
shown in Fig.~\ref{fig5}(e) and Fig.~\ref{fig5}(f). Now one can
among non magnetic phases clearly observe the linear relation
between EFG and $\Delta p$ by comparing the behavior of $V_{zz}$,
Fig.~\ref{fig5}(c), with the behavior of $\Delta p$ shown in
Fig.~\ref{fig5}(d). They behave similar to each other trough non
magnetic phases. This is also the case for the magnetic phases as
well, since the $V_{zz}$ shown in Fig.~\ref{fig5}(e) behaves
similar to $\Delta p$ shown in Fig.~\ref{fig5}(f) through the
magnetic phases. Now time seems apt to intuit that there is an
approximately linear relation between EFG and $DOS(E_F)$, i.e.
$EFG \propto DOS(E_F)$. The relation can be realized, if the
behavior of the total $DOS(E_F)$ shown in Fig.~\ref{fig4}(a) is
compared with the behavior of the $V_{zz}$ shown in
Fig.~\ref{fig5}(c) and Fig.~\ref{fig5}(e). The comparison yields
the result that \textit{the EFG is approximately proportional to
the $DOS(E_F)$}. The result is obtained, because the $V_{zz}$ and
the $DOS(E_F)$ vary similarly, going through all the phases. This
is analogous to the used strategy in sec.~\ref{cv} to realize the
linear relation of $C_V /T = 1/3\pi ^2 k_B^2 DOS(E_F )$. Our
result shows that the proportional constant is negative for the
non magnetic phases, while it is positive for the magnetic phases.
The former constant is negative, because the $V_{zz}$ and
$DOS(E_F)$ inversely vary through non magnetic phases. They,
however, vary directly versus magnetic phases resulting to the
positive constant.

Furthermore, one can also conclude that \textit{the value of
4f-DOS(E$_{F})$ can significantly influence the value of EFG.}
One may confirm this conclusion due to a similarity between the
behavior of 4f $DOS(E_F)$ and total $DOS(E_F)$ shown in
Fig.~\ref{fig4} (a). The foregoing result of $EFG \propto
DOS(E_F)$ ensures that \textit{the EFG can be also proportional
to the 4f $DOS(E_F)$}; i.e. $EFG \propto DOS^{4f}(E_F)$.

The later proportionality between EFG and 4f $DOS(E_F)$ makes more
crucial the method of treatment with 4f-Ce electrons. The later
point describes why the value of EFG at $^{115}In$ site within
the LDA+U calculation is smaller than those obtained within all
the other calculations performed in the lack of LDA+U
interactions. The description can be provided taking the
splitting of 0.43 Ry shown in Fig.~\ref{fig3}(b) into account
between occupied and unoccupied bands due to the LDA+U treatment.
The splitting gives rise to be shifted downwards the 4f DOS from
the vicinity of Fermi level, and as a result to be reduced the
value of 4f $DOS(E_F)$. The later reduction together with the
discussed relation of $EFG \propto DOS^{4f}(E_F)$ provides the
satisfactory description why the EFG is reduced within LDA+U
calculations. All these support our previously concluded result in
sec.~\ref{sub-electr} concerning the important role of 4f
electrons in this compound.

At Ce site due to its cubic point group the EFG's are zero, as
listed in Tab.~\ref{tab1}, for all non magnetic phases and FM
phase. For FM+SOC phase and all the AFM phases, however, our
result shows non zero EFG values at this site. The non zero values
for the EFG's originate from the fact that SOC or magnetic
ordering can give rise to a little bit deviation from cubic
symmetry. One expects that the small deviation gives rise to
small EFG's. Our result, in Tab.~\ref{tab1}, confirms the later
point apart from the last phase of AFM(111)+SOC+LDA+U. For the
later AFM(111)+SOC+LDA+U phase the EFG at Ce site is also
practically small, but it is 2 order of magnitude larger than the
other phases. To find the source of such a discrepancy, we follow
our last strategy looking at the behavior of EFG and anisotropy
functions through corresponded phases. The calculated EFG at Ce
site can be compared with the $\Delta p$ versus magnetic phases
(apart from FM phase for which EFG is exactly zero) using
Fig.~\ref{fig6} (a) and (b). The comparison does not show similar
behavior for the EFG and $\Delta p$ along the magnetic phases. To
find the reason why they do not show similar behavior, now we
come back to the mentioned point that $|\Delta d (E_F)|>|\Delta p
(E_F)|$ at the Ce site; see Tab.~\ref{tab2}. One first suspects,
due to the larger value of $|\Delta d (E_F)|$ than $|\Delta p
(E_F)|$, that the behavior of EFG might be similar to the
behavior of $\Delta d (E_F)$. The behaviors of EFG and $\Delta d
(E_F)$ are compared in Fig.~\ref{fig6} (a) and (c). However, they
also do not show similar behavior. Therefore, we could not
reproduce the behavior of the EFG using either $\Delta p (E_F)$
or $\Delta d (E_F)$. For more realization, we have decomposed the
EFG at Ce site into its valance contributions. The results are
shown in Figs.~\ref{fig7} for spin up and spin down. In the
Figs.~\ref{fig7}, we have inset a figure to show sum of up and
down spins for each of the valance contributions individually.
The result, comparing Fig.~\ref{fig7} (a) and (b), shows that
contributions of d-states to the EFG is one order of magnitude
less than contributions of p-states. This is in the case that, as
expressed before, $|\Delta d(E_F)|$ is one order of magnitude
larger than $|\Delta p(E_F)|$. \textit{Therefore, even for the
case of $|\Delta d(E_F)| > |\Delta p(E_F)|$, contributions to the
EFG originating from p-states compared to the d-states dominate.}
This result is in complete accord with the calculations performed
in Ref. \onlinecite{Blaha1988} for transition metals. However, a
different and important result can be emerged from comparing
Fig.~\ref{fig7} (c) with Fig.~\ref{fig6} (a). \textit{The result
is this that the EFG mainly originates from f-states at Ce site.}
We have added all the valance contributions to the EFG in
Fig.~\ref{fig7} (d). The result shows that the behavior of EFG
shown in Fig.~\ref{fig6} (a) is similar to total valance EFG
contributions shown in Fig.~\ref{fig7} (c). Lattice contributions
to the EFG also cannot change the similarity, because they are
negligible as listed in Tab.~\ref{tab1}. Therefore
\textit{contributions to the EFG originating from f states
dominate at Ce site.} For sure in Fig.~\ref{fig7} (e) we have
subtracted contributions to the EFG of f-states from total
valance EFG contributions. The result, Fig.~\ref{fig7} (e),
represents the behavior of p-anisotropy function shown in
Fig.~\ref{fig6} (b). Such a representation reconfirms our last
conclusion.


\section{Conclusion}
\label{conclu}

We have investigated the variations of electric field gradient
(EFG's) and their valance contributions as well as their
anisotropy functions at both In and Ce sites in the $CeIn_3$ for a
variety of circumstances. For each of the circumstances, density
of states at Fermi level ($DOS(E_F)$) is calculated. We have
found that comparing the behavior of the EFG with the behavior of
$DOS(E_F)$ versus the applied circumstances may be a useful
strategy to emerge physical properties. We have found within such
a strategy that the main component of the electric field gradient,
$V_{zz}$, is approximately proportional to the value of total
density of states at Fermi level ($DOS(E_F)$) as well as
$4f-DOS(E_F)$. Despite anisotropy function of d-states is larger
than the one of p-states at Ce site, contributions to the EFG
originating from p-states dominate compared to d-states. This is
in the case that, however, the EFG's are dominated at the
$^{140}Ce$ site by the 4f-states, while they are as usual
dominated by p-states at the In site. The result shows that 4f Ce
states are hybridized with conduction bands and play an important
role which must not be ignored. Thus we have within our electronic
structure calculations predicted that neither band-like treatment,
nor open-core treatment can provide satisfactory results for this
case. Our LDA+U calculations are in better agreement with
experiment indicating the fact that correlations among 4f
electrons are taken more properly into account. The correlations
among 4f electrons influence not only 4f Ce states directly, but
also 5p Ce semicore states indirectly in this compound.

\acknowledgments This work has been performed based on the
research project number 821235, University of Isfahan (UI),
Isfahan, Iran.

\end{document}